\documentclass[acmsmall,nonacm]{acmart}

\usepackage{makecell}
\usepackage{bm}
\usepackage{enumitem}
\setlist{nolistsep}
\usepackage{multirow}
\usepackage{float}
\usepackage{tabularx}
\usepackage{pifont}
\usepackage{booktabs}
\usepackage{caption}
\usepackage{colortbl}
\usepackage[subrefformat=parens,skip=0pt]{subcaption}
\usepackage{overpic}
\usepackage{marginnote}
\usepackage[export]{adjustbox}
\usepackage[detect-all]{siunitx}
\usepackage{xspace}

\newcommand\ie{i.\,e.\xspace}
\newcommand\eg{e.\,g.\xspace}

\newcommand\US{U.S.\xspace}

\newcommand{\var}[1]{\mathit{#1}}

\def\sym#1{\ifmmode^{#1}\else\(^{#1}\)\fi}

\usepackage{soul}
\usepackage{ifthen}
\newif\ifcommenton
\commentontrue
\AtBeginDocument{%
  }

\begin{document}
\title{Changes in Sentiments and User Engagement for 2024 U.S. Presidential Candidates After Biden's Withdrawal: An Analysis of TikTok Videos}

\author{Yuwei Chuai}
\affiliation{
  \institution{University of Luxembourg}
  \country{Luxembourg}}
\email{yuwei.chuai@uni.lu}
\orcid{0000-0001-6181-7311}

\author{Gabriele Lenzini}
\affiliation{
  \institution{University of Luxembourg}
  \country{Luxembourg}}
\email{gabriele.lenzini@uni.lu}
\orcid{0000-0001-8229-3270}

\begin{abstract}
The 2024 \US presidential election has sparked widespread online discussions about the presidential candidates. Joe Biden's withdrawal from the race and Kamala Harris's subsequent entry as the Democratic candidate likely alter the dynamics of these online discussions; yet, this hypothesis requires evidence. Here, we study how sentiments and user engagement in social media posts mentioning presidential candidates change after Biden's withdrawal. Our analysis is based on $N=$ \num{680609} TikTok videos that have accumulated over 4 billion views, with more than 23 million comments, 31 million shares, and 335 million likes from November 1, 2023, to October 6, 2024. We find that: (i) Before Biden's withdrawal, video posts mentioning the Republican candidate (Donald Trump) have higher positive sentiment and lower negative sentiment compared to those mentioning the Democratic candidate (Joe Biden). (ii) Following Biden's withdrawal, positive sentiment in video posts mentioning the Democratic candidate (Kamala Harris) increases by 46.8\%, while negative sentiment decreases by 52.0\%. (iii) Regarding user engagement, before Biden's withdrawal, video posts mentioning the Democratic candidate have 64.9\% higher odds of being shared and 39.5\% higher odds of receiving likes compared to posts mentioning the Republican candidate, with similar odds of receiving comments. (iv) After Biden's withdrawal, the odds of being shared increase by 53.3\%, and the odds of receiving likes increase by 77.4\% in both video posts mentioning the Democratic candidate and video posts mentioning the Republican candidate. Our findings offer insights into how sentiments and user engagement in online posts about the 2024 \US presidential candidates shift following Biden's dropping out from the presidential race.
\end{abstract}

\maketitle

% \newpage
\section{Introduction}
Social media platforms, such as TikTok, Facebook, and X (formerly Twitter), have become major tools for news consumption and user engagement, playing a crucial role in shaping user opinions and perceptions of real-world events, \eg, political elections. According to the report from TikTok, more than 170 million users -- about half of the \US population -- consume content on its video-sharing platform, and the platform is actively taking measures to protect election integrity in 2024.\footnote{\url{https://newsroom.tiktok.com/en-us/continuing-to-protect-the-integrity-of-tiktok-through-the-us-elections}} Currently, the 2024 \US presidential race has been marked by key events, including the first presidential debate between Republican candidate Donald Trump and Democratic candidate Joe Biden on June 27, 2024, Joe Biden's withdrawal from the race on July 21, 2024, and the second presidential debate between Donald Trump and Kamala Harris (Biden's replacement for Democratic candidate) on September 10, 2024. This paper analyzes sentiments and user engagement in TikTok video posts mentioning presidential candidates during those events. Specifically, we explore whether there are significant changes in sentiments and user engagement in video posts mentioning presidential candidates following Biden's withdrawal.

\section{Data Collection}
We use TikTok's official API for academic research\footnote{The research API on TikTok is open for applications for researchers in the \US and Europe via \url{https://developers.tiktok.com/products/research-api/}} to collect video posts that mention 2024 \US presidential candidates, including Donald Trump (Republican candidate), Joe Biden (Democratic candidate), and Kamala Harris (Democratic candidate). Using the video query request, we search for posts that are located in the region of \US and mention only one of the three candidates. For example, video posts for Donald Trump do not mention Joe Biden or Kamala Harris. 

The API request limit is \num{1000} per day, and each video query request can return up to \num{100} records. Given this limitation, we repeat the same request 20 times per day for each candidate, with each request returning a maximum of 100 random videos. Each video object includes its unique ID, creation time, username, video description, voice-to-text transcription, view count, comment count, share count, and like count.

On July 21, 2024, Biden dropped out of the presidential race and was replaced by Harris as the Democratic candidate. Therefore, we collect video posts mentioning ``T[t]rump'' or ``B[b]iden'' before Biden's withdrawal and video posts mentioning ``T[t]rump'' or ``H[h]arris'' afterwards. As a result, we gathered \num{680609} video posts created between November 1, 2023, and October 6, 2024. These video posts have accumulated over 4 billion views, with more than 23 million comments, 31 million shares, and 335 million likes. 

\section{Methods}
\subsection{Calculation of Sentiments}
We combine the video description and voice-to-text transcription to calculate positive and negative sentiments for each video post, using a state-of-the-art machine-learning model.\footnote{\url{https://huggingface.co/cardiffnlp/twitter-roberta-base-sentiment-latest}} Each video post is assigned a score of positive sentiment ($\in [0,1]$), a score of neutral sentiment ($\in [0,1]$), a score of negative sentiment ($\in [0,1]$), with the three scores summing to 1. We focus only on the positive and negative sentiments in the later analysis.

\subsection{Definition of Engagement Rates}
Engagement with video posts includes comments, shares, and likes. We define engagement rates as follows:
\begin{itemize}[leftmargin=*]
    \item \textbf{Comment rate}: The ratio of comment count to view count.
    \item \textbf{Share rate}: The ratio of share count to view count.
    \item \textbf{Like rate}: The ratio of like count to view count.
\end{itemize}
To evaluate the engagement rates, we restrict our analysis to video posts that have been viewed at least once, and where the comment count, share count, and like count are less than or equal to the corresponding view count. This results in \num{656115} video posts, accounting for 96.4\% of the total video posts.

\subsection{Empirical Models}
We consider the video posts created between the first debate and the second debate and specify a linear regression model to estimate the changes in sentiments following Biden's withdrawal. Additionally, we use a binomial regression model to estimate the effects of sentiments and Biden's withdrawal on engagement rates. To this end, we define the following variables:
\begin{itemize}[leftmargin=*]
    \item $\var{Positive}$: A continuous variable from 0 to 1 that indicates the probability of positive sentiment in a video post.
    \item $\var{Negative}$: A continuous variable from 0 to 1 that indicates the probability of negative sentiment in a video post.
    \item $\var{ViewCount}$: A count variable that indicates the number of views for a video post.
    \item $\var{CommentCount}$: A count variable that indicates the number of comments for a video post.
    \item $\var{ShareCount}$: A count variable that indicates the number of shares for a video post.
    \item $\var{LikeCount}$: A count variable that indicates the number of likes for a video post.
    \item $\var{Democratic}$: A binary variable that is equal to 1 if the video post mentions Democratic candidate (\ie, Biden or Harris) and 0 if the video post mentions Republican candidate (\ie Trump).
    \item $\var{BidenDropOut}$: A binary variable that equals 1 if the video post is created after Biden's withdrawal, and 0 otherwise.
    \item $\var{DaysFromDropOut}$: A count variable that indicates the number of days from Biden's withdrawal.
\end{itemize}

Subsequently, we specify the following linear regression model:
\begin{equation}
 \begin{aligned}
    \var{Sentiment_{i}} & = \beta_{0} + \beta_{1}\var{Democratic_{i}} + \beta_{2}\var{BidenDropOut_{i}} + \beta_{3}\var{Democratic_{i}} \times \var{BidenDropOut_{i}} \\ 
    &+ \beta_{4}\var{DaysFromDropOut_{i}} + \epsilon_{i},
\end{aligned}
\end{equation}
where $\var{Sentiment}$ denotes the dependent variable of $\var{Positive}$ or $\var{Negative}$, $\var{\beta}_{0}$ is the intercept, and $\epsilon$ is the residual. $\var{DaysFromDropOut}$ is $z$-standardized. $\var{\beta}_{1}$ to $\var{\beta}_{4}$ are coefficient estimates for the independent variables. $\var{\beta}_{1}$ represents the average difference in sentiments between video posts mentioning Democratic candidate and video posts mentioning Republican candidate before Biden's withdrawal. $\var{\beta}_{2}$ represents the shared change in sentiments between video posts mentioning Democratic candidate and video posts mentioning Republican candidate after Biden's withdrawal. $\var{\beta}_{3}$ represents the unique change in sentiments in video posts mentioning Democratic candidate following Biden's withdrawal. Additionally, we assume that comments, shares, and likes follow binomial distributions:
\begin{equation}
\begin{aligned}
&\var{CommentCount_{i}} \sim binomial(ViewCount_{i}, \theta_{i}^{c}),\\
&\var{ShareCount_{i}} \sim binomial(ViewCount_{i}, \theta_{i}^{s}),\\
&\var{LikeCount_{i}} \sim binomial(ViewCount_{i}, \theta_{i}^{l}),\\
\end{aligned}
\end{equation}
where $\var{\theta_{i}} \in [0,1]$ represents the probability that a user will comment, share, or like a video post, which corresponds to the engagement rates. Using a $\var{logit}$ link function, we specify the following binomial regression model:
\begin{equation}
\begin{aligned}
    logit(\theta_{i}) & = \beta_{0} + \beta_{1}\var{Democratic_{i}} + \beta_{2}\var{BidenDropOut_{i}} + \beta_{3}\var{Democratic_{i}} \times \var{BidenDropOut_{i}}\\ 
    &+ \beta_{4}\var{Positive_{i}} + \beta_{5}\var{Negative_{i}} + \beta_{6}\var{DaysFromDropOut_{i}},
\end{aligned}
\end{equation}
where $\var{Positive}$, $\var{Negative}$, and $\var{DaysFromDropOut}$ are $z$-standardized. $\var{\beta}_{1}$ represents the average difference in engagement rates between video posts mentioning Democratic candidate and video posts mentioning Republican candidate before Biden's withdrawal. $\var{\beta}_{2}$ represents the shared change in engagement rates between video posts mentioning Democratic candidate and video posts mentioning Republican candidate after Biden's withdrawal. $\var{\beta}_{3}$ represents the unique change in engagement rates in video posts mentioning Democratic candidate following Biden's withdrawal.

\section{Descriptive Analysis}
\subsection{Sentiments}
\begin{figure}
\centering
\begin{subfigure}{\textwidth}
\includegraphics[width=\textwidth]{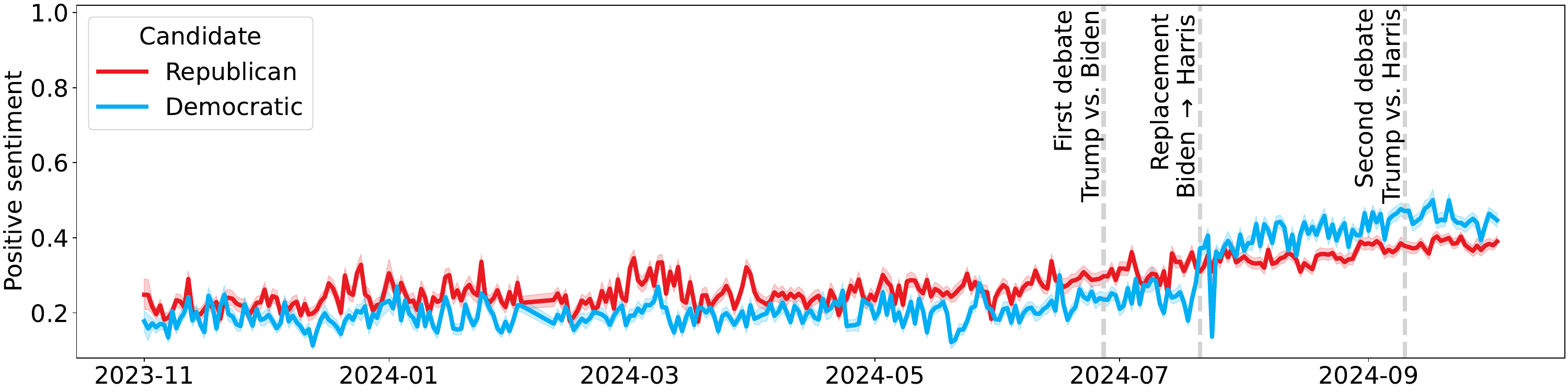}
\caption{}
\label{fig:positive}
\end{subfigure}

\begin{subfigure}{\textwidth}
\includegraphics[width=\textwidth]{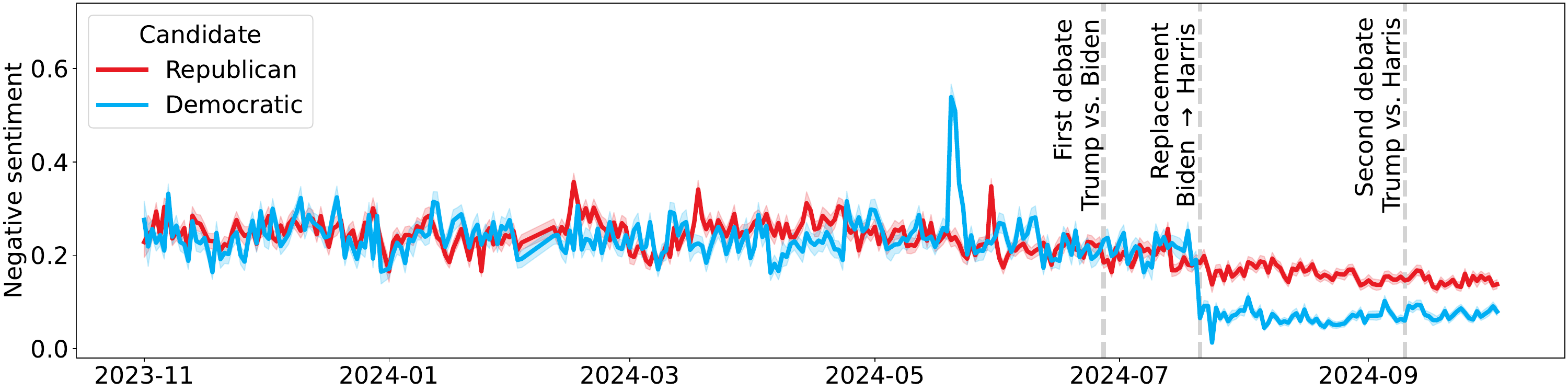}
\caption{}
\label{fig:negative}
\end{subfigure}
\caption{The daily averages of positive and negative sentiments in TikTok video posts mentioning Republican and Democratic presidential candidates from November 1, 2023, to October 6, 2024. \subref{fig:positive} shows positive sentiment, and \subref{fig:negative} shows negative sentiment. The error bands represent 95\% CIs. The first (left) vertical dashed line indicates the first 2024 \US presidential debate between Donald Trump and Joe Biden, the second (middle) dashed line indicates Biden's withdrawal from the race, and the third (right) dashed line marks the second presidential debate between Donald Trump and Kamala Harris.}
\label{fig:sentiments}
\end{figure}

We start by analyzing the overall changes in sentiments and user engagement rates in video posts mentioning presidential candidates from November 1, 2023, to October 6, 2024. Fig. \ref{fig:sentiments} shows the daily averages of positive and negative sentiments in TikTok video posts mentioning Republican and Democratic candidates. We find a sharp increase in positive sentiment (Fig. \ref{fig:positive}) and a sharp decrease in negative sentiment (Fig. \ref{fig:negative}) in video posts mentioning Democratic candidate following Biden's withdrawal from the presidential race, while the positive and negative sentiments remain relatively stable following the first and second presidential debates.

\subsection{Engagement rates}
\begin{figure}
\centering
\begin{subfigure}{\textwidth}
\includegraphics[width=\textwidth]{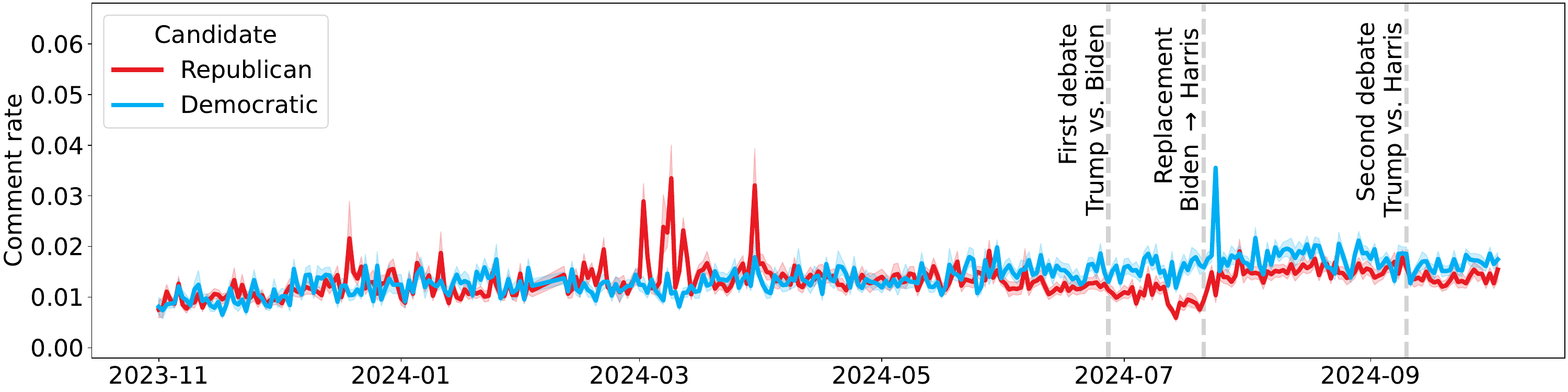}
\caption{}
\label{fig:comment_rate}
\end{subfigure}

\begin{subfigure}{\textwidth}
\includegraphics[width=\textwidth]{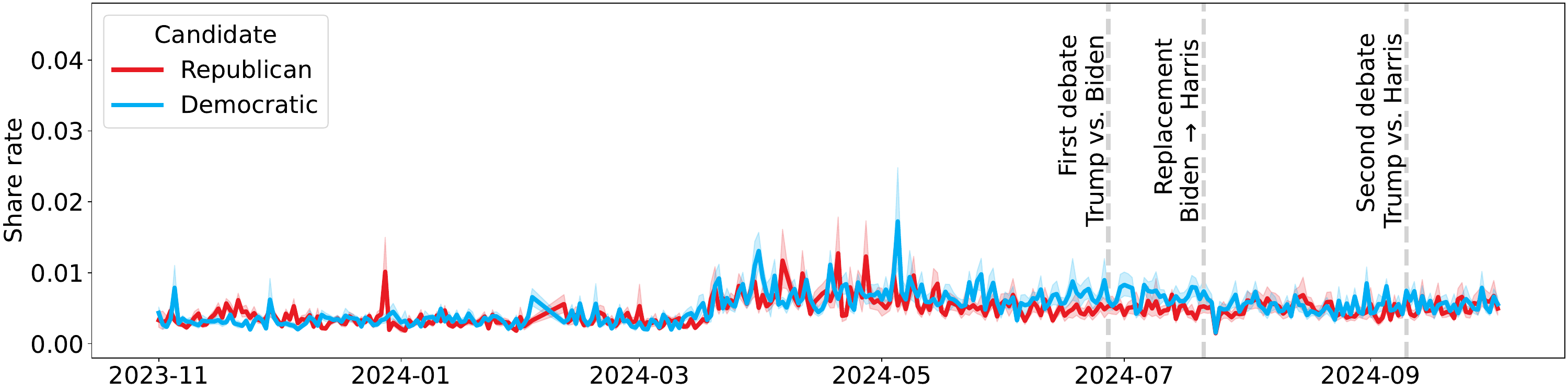}
\caption{}
\label{fig:share_rate}
\end{subfigure}

\begin{subfigure}{\textwidth}
\includegraphics[width=\textwidth]{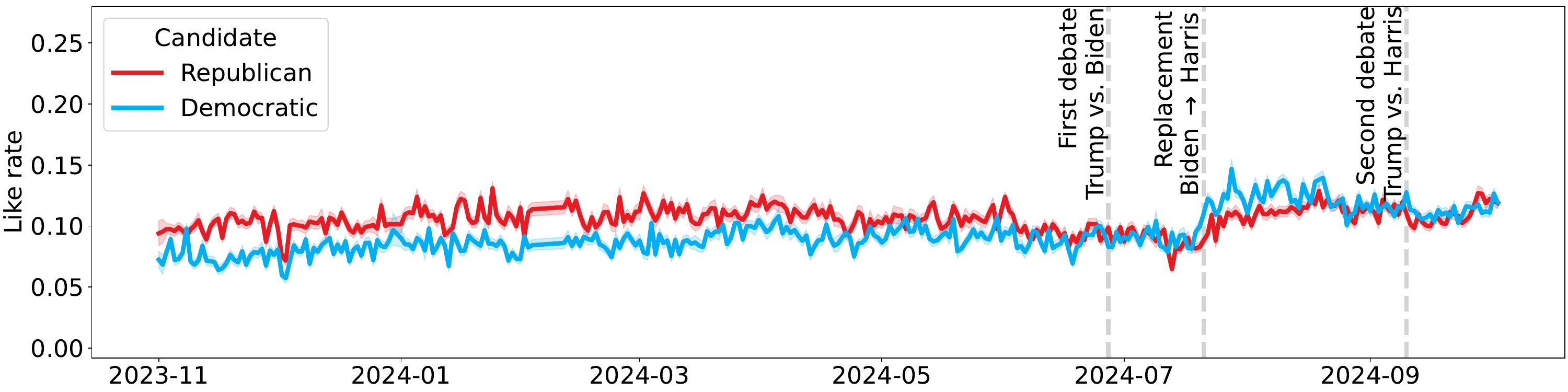}
\caption{}
\label{fig:like_rate}
\end{subfigure}
\caption{The daily averages of engagement rates in TikTok video posts mentioning Republican and Democratic presidential candidates from November 1, 2023, to October 6, 2024.  \subref{fig:comment_rate} shows the comment rate, \subref{fig:share_rate} shows the share rate, and \subref{fig:like_rate} shows the like rate. The error bands represent 95\% CIs. The first (left) vertical dashed line indicates the first 2024 \US presidential debate between Donald Trump and Joe Biden, the second (middle) dashed line indicates Biden's withdrawal from the race, and the third (right) dashed line marks the second presidential debate between Donald Trump and Kamala Harris.}
\label{fig:engagement_rates}
\end{figure}

Fig. \ref{fig:engagement_rates} shows the daily averages of engagement rates in TikTok videos mentioning Republican and Democratic candidates from November 1, 2023 to October 6, 2024. We find that:
\begin{itemize}[leftmargin=*]
    \item \textbf{Comment rate} in Fig. \ref{fig:comment_rate}
        \begin{itemize}
            \item Overall, the comment rate in video posts mentioning Republican candidate is similar to that in video posts mentioning Democratic candidate.
            \item A peak of comment rate is observed in video posts mentioning Democratic candidate right after Biden's withdrawal from the race.
            \item The comment rate in video posts mentioning Republican candidate shows several peaks around March of 2024. Additionally, it decreases after the first debate but returns to its original level following Biden's withdrawal from the race.
        \end{itemize}
    \item \textbf{Share rate} in Fig. \ref{fig:share_rate}
    \begin{itemize}
        \item Overall, the share rate in video posts mentioning Republican candidate shows a similar trend to that in video posts mentioning Democratic candidate, regardless of the debates or Biden's withdrawal.
        \item The share rate in video posts either mentioning Republican candidate or Democratic candidate increases after March of 2024.
    \end{itemize}
    \item \textbf{Like rate} in Fig. \ref{fig:like_rate}
    \begin{itemize}
        \item Video posts mentioning Republican candidate show a higher like rate than those mentioning Democratic candidate before the first presidential debate. 
        \item After the first presidential debate, the like rates for video posts mentioning the Republican and Democratic candidates overlap, and both increase following Biden's withdrawal from the race.
    \end{itemize}
\end{itemize}

\subsection{Summary of Descriptive Analysis}
In summary, we observe a significant increase in positive sentiment and a significant decrease in negative sentiment in video posts mentioning Democratic candidate following Biden's withdrawal from the race. However, in terms of the engagement rates, only the like rate increases significantly following Biden's withdrawal, and this is shown in both video posts mentioning Republican candidate and video posts mentioning Democratic candidate. Of note, the changes in positive and negative sentiments following Biden's withdrawal are likely to affect engagement rates. Therefore, the effects of Biden's withdrawal on the engagement rates could be confounded with the effects of sentiments. To address this, we employ regression models to quantitatively analyze the changes in sentiments following Biden's withdrawal and estimate the effects of both sentiments and Biden's withdrawal on engagement rates.

\section{Empirical Analysis}
\subsection{Sentiments}
\begin{figure}
\centering
\begin{subfigure}{0.32\textwidth}
\includegraphics[width=\textwidth]{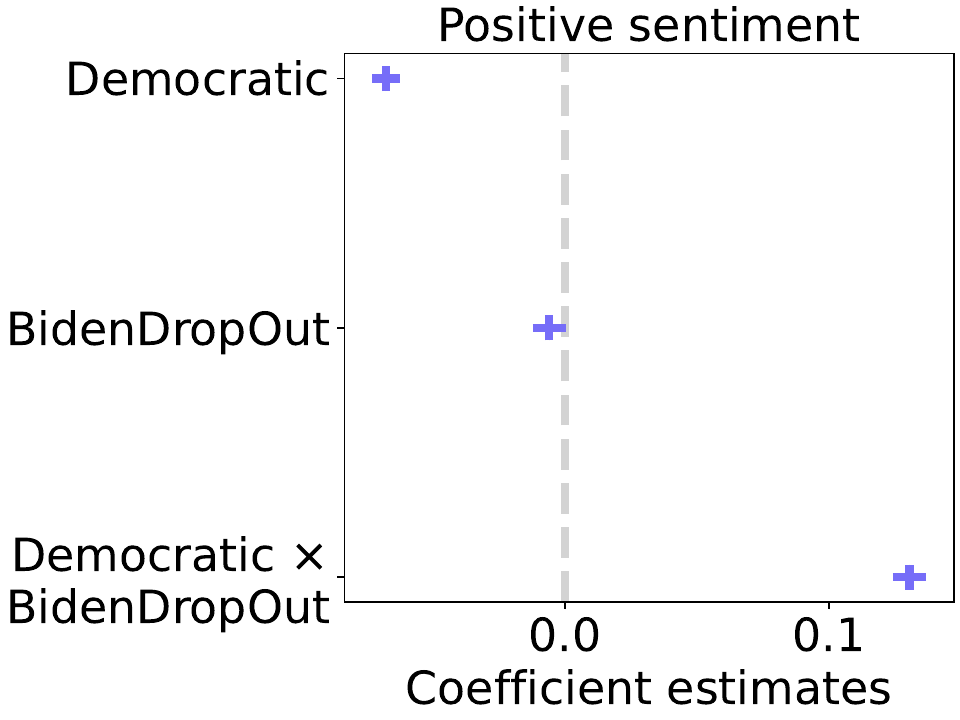}
\caption{}
\label{fig:positive_coefs}
\end{subfigure}
\hspace{1cm}
\begin{subfigure}{0.32\textwidth}
\includegraphics[width=\textwidth]{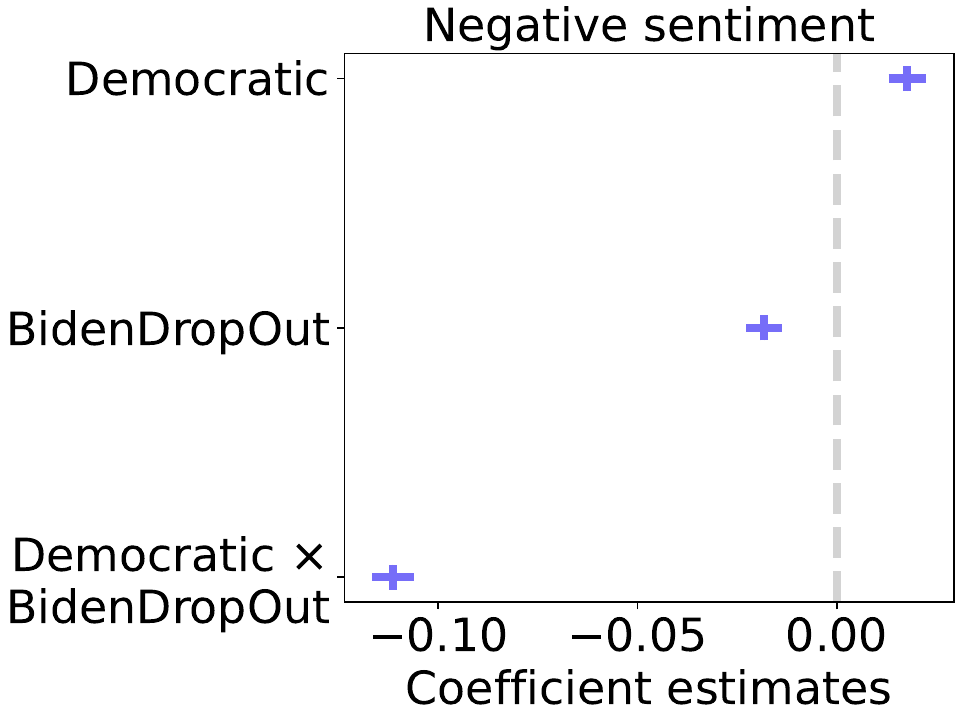}
\caption{}
\label{fig:negative_coefs}
\end{subfigure}
\caption{The estimated coefficients for the independent variables -- $\var{Democratic}$ (1 indicates Democratic candidate; 0 indicates Republican candidate), $\var{BidenDropOut}$, $\var{Democratic} \times \var{BidenDropOut}$. The independent variable $\var{DaysFromDropOut}$ is included during estimation but omitted in the visualization for better readability. Shown are mean values with error bars representing 95\% CIs. The dependent variables are \subref{fig:positive_coefs} positive sentiment and \subref{fig:negative_coefs} negative sentiment in TikTok video posts mentioning 2024 \US presidential candidates, respectively.}
\label{fig:sentiment_coefs}
\end{figure}

We use video posts created between the first and second presidential debates to estimate the changes in sentiments following Biden's withdrawal. Fig. \ref{fig:sentiment_coefs} shows the coefficient estimates for predicting positive and negative sentiments. We find that:
\begin{itemize}[leftmargin=*]
    \item \textbf{Positive sentiment} in Fig. \ref{fig:positive_coefs}
    \begin{itemize}
        \item The coefficient estimate of $\var{Democratic}$ is significantly negative ($\var{coef.}=-0.068$, $p<0.001$; 95\% CI: $[-0.073, -0.063]$). This indicates that, before Biden's withdrawal, the positive sentiment in video posts mentioning Democratic candidate is 0.068 lower than that in video posts mentioning Republican candidate. 
        \item The coefficient estimate of $\var{BidenDropOut}$ is not statistically significant. However, the coefficient estimate of $\var{Democratic} \times \var{BidenDropOut}$ is significantly positive ($\var{coef.}=0.130$, $p<0.001$; 95\% CI: $[0.124, 0.137]$). This indicates that positive sentiment in video posts mentioning Democratic candidate increases by 0.130 after Biden's withdrawal, while there is no significant change in the positive sentiment in video posts mentioning Republican candidate. 
        \item Using the positive sentiment in video posts mentioning Democratic candidate before Biden's withdrawal as a baseline, the positive sentiment in video posts mentioning Democratic candidate increases by 46.8\% following Biden's withdrawal.
    \end{itemize}
    \item \textbf{Negative sentiment} in Fig. \ref{fig:negative_coefs}
    \begin{itemize}
        \item The coefficient estimate of $\var{Democratic}$ is significantly positive ($\var{coef.}=0.018$, $p<0.001$; 95\% CI: $[0.013, 0.022]$). This indicates that, before Biden's withdrawal, negative sentiment in video posts mentioning Democratic candidate is 0.018 higher than that in video posts mentioning Republican candidate.
        \item The coefficient estimate of $\var{BidenDropOut}$ is significantly negative ($\var{coef.}=-0.018$, $p<0.001$; 95\% CI: $[-0.023, -0.014]$). This indicates that negative sentiment in both video posts mentioning Democratic candidate and video posts mentioning Republican candidate decreases by 0.018 after Biden's withdrawal. 
        \item Moreover, the coefficient estimate of $\var{Democratic} \times \var{BidenDropOut}$ is significantly negative ($\var{coef.}=-0.111$, $p<0.001$; 95\% CI: $[-0.116, -0.106]$). This indicates that negative sentiment in video posts mentioning Democratic candidate additionally decreases by 0.111 after Biden's withdrawal.
        \item Using the negative sentiment in video posts mentioning Democratic candidate before Biden's withdrawal as a baseline, negative sentiment in video posts mentioning Democratic candidate additionally decreases by 52.0\% following Biden's withdrawal, beyond the shared decrease observed in video posts mentioning Republican candidate.
    \end{itemize}
\end{itemize}

\subsection{Engagement rates}
\begin{figure}
\centering
\begin{subfigure}{0.32\textwidth}
\includegraphics[width=\textwidth]{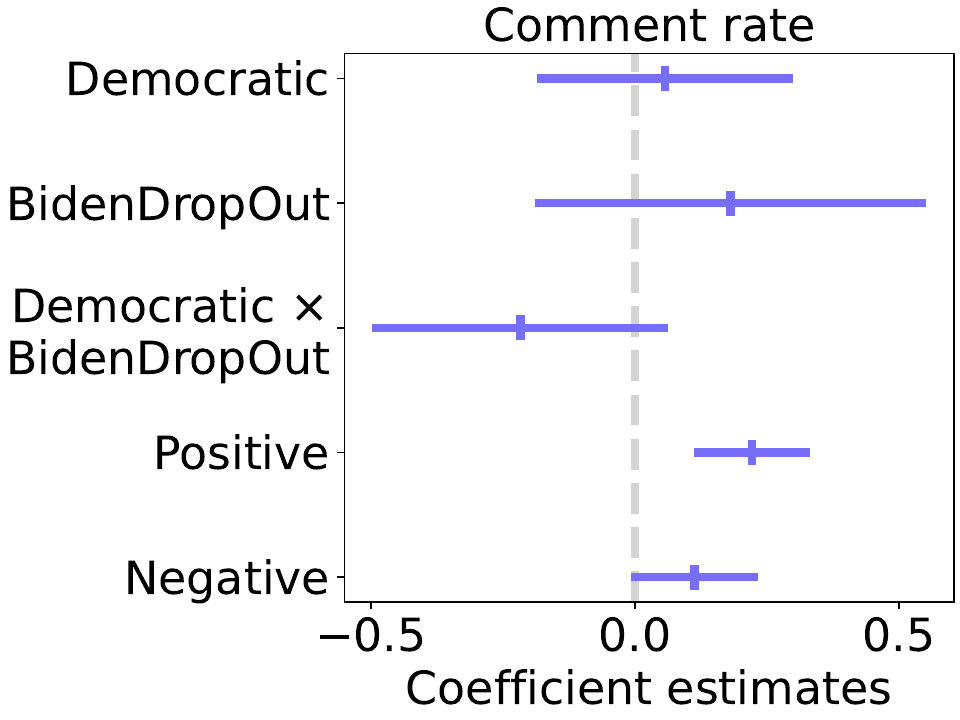}
\caption{}
\label{fig:comment_rate_coefs}
\end{subfigure}
\hfill
\begin{subfigure}{0.32\textwidth}
\includegraphics[width=\textwidth]{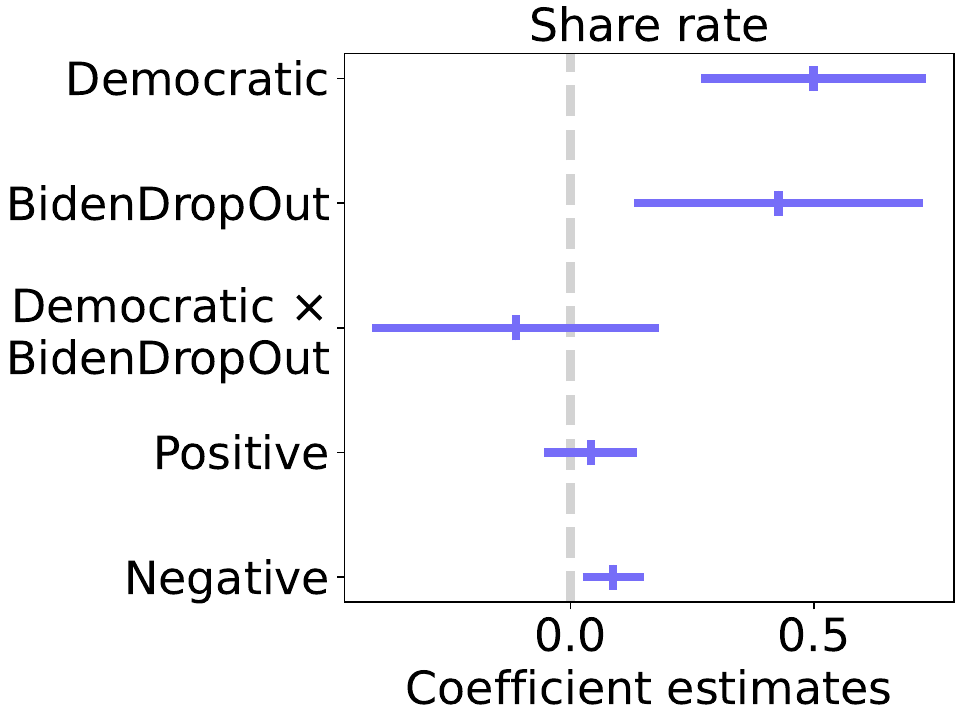}
\caption{}
\label{fig:share_rate_coefs}
\end{subfigure}
\hfill
\begin{subfigure}{0.32\textwidth}
\includegraphics[width=\textwidth]{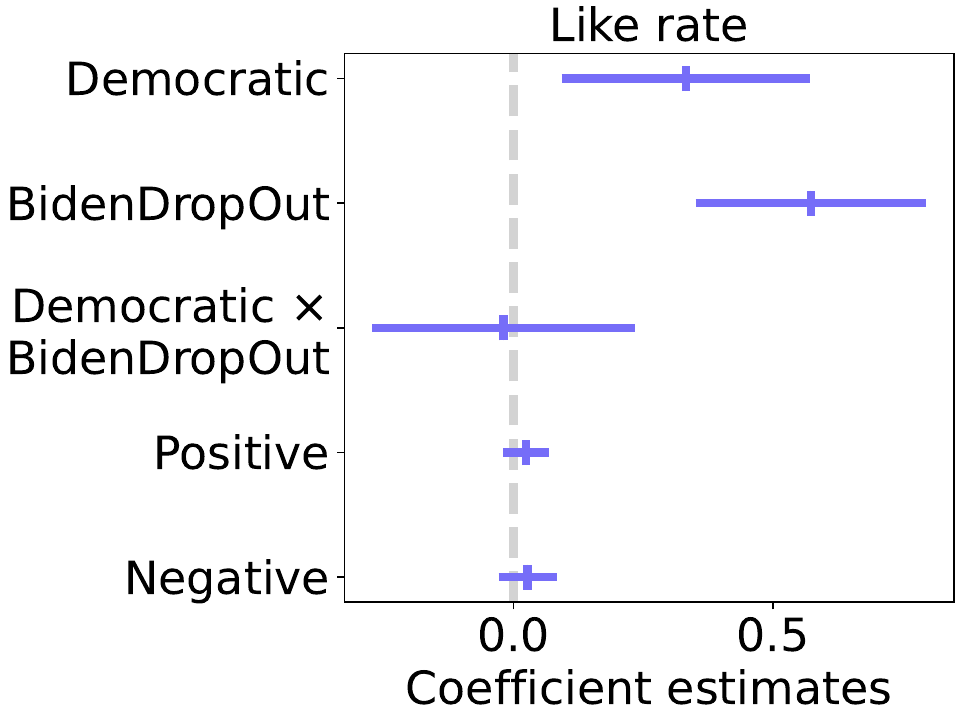}
\caption{}
\label{fig:like_rate_coefs}
\end{subfigure}
\caption{The estimated coefficients for the independent variables -- $\var{Democratic}$ (1 indicates Democratic candidate; 0 indicates Republican candidate), $\var{BidenDropOut}$, $\var{Democratic} \times \var{BidenDropOut}$, $\var{Positive}$, and $\var{Negative}$. The independent variable $\var{DaysFromDropOut}$ is included during estimation but omitted in the visualization for better readability. Shown are mean values with error bars representing 95\% CIs. The dependent variables are \subref{fig:comment_rate_coefs} comment rate, \subref{fig:share_rate_coefs} share rate, and \subref{fig:like_rate_coefs} like rate in TikTok video posts mentioning 2024 \US presidential candidates, respectively.}
\label{fig:engagement_rate_coefs}
\end{figure}

Fig. \ref{fig:engagement_rate_coefs} shows the coefficient estimates for predicting engagement rates. We find that:
\begin{itemize}[leftmargin=*]
    \item \textbf{Comment rate} in Fig. \ref{fig:comment_rate_coefs}
    \begin{itemize}
        \item The coefficient estimate of $\var{Democratic}$ is not statistically significant. This indicates that, before Biden's withdrawal, the comment rates in video posts mentioning Democratic candidate and video posts mentioning Republican candidate have no significant difference.
        \item The coefficient estimates for both $\var{BidenDropOut}$ and $\var{Democratic} \times \var{BidenDropOut}$ are not statistically significant. This indicates that Biden's withdrawal has no significant effect on the comment rate in either video posts mentioning Democratic candidate or video posts mentioning Republican candidate.
        \item The coefficient estimate of $\var{Positive}$ is significantly positive ($\var{coef.}=0.221$, $p<0.001$; 95\% CI: $[0.114, 0.328]$). This indicates that a one-standard-deviation increase in positive sentiment is associated with 24.7\% higher odds of receiving comments (95\% CI: $[0.121, 0.388]$).
    \end{itemize}
    \item \textbf{Share rate} in Fig. \ref{fig:share_rate_coefs}
    \begin{itemize}
        \item The coefficient estimate of $\var{Democratic}$ is significantly positive ($\var{coef.}=0.500$, $p<0.001$; 95\% CI: $[0.273, 0.726]$). This indicates that, before Biden's withdrawal, video posts mentioning Democratic candidate have 64.9\% (95\% CI: $[0.314, 1.067]$) higher odds of being shared compared to video posts mentioning Republican candidate.
        \item The coefficient estimate of $\var{BidenDropOut}$ is significantly positive ($\var{coef.}=0.427$, $p<0.01$; 95\% CI: $[0.136, 0.718]$). This indicates that Biden's withdrawal is associated with 53.3\% (95\% CI: $[0.146, 1.050]$) higher odds of being shared for both video posts mentioning Democratic candidate and video posts mentioning Republican candidate.
        \item The coefficient estimate of $\var{Democratic} \times \var{BidenDropOut}$ is not statistically significant. This suggests that, following Biden's withdrawal, the change in the share rate in video posts mentioning Democratic candidate has no significant difference from that in video posts mentioning Republican candidate.
        \item The coefficient estimate of $\var{Negative}$ is significantly positive ($\var{coef.}=0.088$, $p<0.01$; 95\% CI: $[0.026, 0.149]$). This indicates that a one-standard-deviation increase in negative sentiment is associated with 9.2\% higher odds of being shared (95\% CI: $[0.026, 0.161]$).
    \end{itemize}
    \item \textbf{Like rate} in Fig. \ref{fig:like_rate_coefs}
    \begin{itemize}
        \item The coefficient estimate of $\var{Democratic}$ is significantly positive ($\var{coef.}=0.333$, $p<0.01$; 95\% CI: $[0.098, 0.567]$). This indicates that, before Biden's withdrawal, video posts mentioning Democratic candidate have 39.5\% (95\% CI: $[0.103, 0.763]$) higher odds of receiving likes compared to video posts mentioning Republican candidate.
        \item The coefficient estimate of $\var{BidenDropOut}$ is significantly positive ($\var{coef.}=0.573$, $p<0.001$; 95\% CI: $[0.355, 0.791]$). This indicates that Biden's withdrawal is associated with 77.4\% (95\% CI: $[0.426, 1.206]$) higher odds of receiving a like for both video posts mentioning Democratic candidate and video posts mentioning Republican candidate.
        \item The coefficient estimate of $\var{Democratic} \times \var{BidenDropOut}$ is not statistically significant. This suggests that, after Biden's withdrawal, the change in the like rate in video posts mentioning Democratic candidate has no significant difference from that in video posts mentioning Republican candidate.
        \item The coefficient estimates of $\var{Positive}$ and $\var{Negative}$ are not statistically significant. This suggests that the like rate has no significant association with sentiments.
    \end{itemize}
\end{itemize}

\subsection{Summary of Empirical Analysis}
Taken together, we find that, before Biden's withdrawal, video posts mentioning Republican candidate have higher positive sentiment and lower negative sentiment compared to video posts mentioning Democratic candidate. However, following Biden's withdrawal, the positive sentiment in video posts mentioning Democratic candidate increases by 46.8\%, while the negative sentiment decreases by 52.0\%. In terms of user engagement, before Biden's withdrawal, video posts mentioning Democratic candidate have 64.9\% higher odds of being shared and 39.5\% higher odds of receiving likes compared to video posts mentioning Republican candidate, with similar comment rates. Additionally, following Biden's withdrawal, the odds of being shared increases by 53.3\%, and the odds of receiving likes increases by 77.4\% in both video posts mentioning Democratic candidate and video posts mentioning Republican candidate. 

Notably, our findings show that sentiments have significant effects on the comment rate and share rate. Specifically, positive sentiment is positively associated with the comment rate, and negative sentiment is positively associated with the share rate. Therefore, the decrease in negative sentiment after Biden's withdrawal (Fig. \ref{fig:negative}) suggests a potential decrease in the share rate. The effect of Biden's withdrawal on the share rate may be confounded by the effect of negative sentiment, which could partially explain why we do not observe a significant change in the share rate following Biden's withdrawal in the descriptive analysis (Fig. \ref{fig:share_rate}). In contrast, sentiments have no significant effects on the like rate, which helps explain that we can observe a significant increase in the like rate following Biden's withdrawal in the descriptive analysis (Fig. \ref{fig:like_rate}).

\section{Conclusion}
In this paper, we conduct a large-scale analysis of TikTok video posts in terms of sentiments and user engagement. We find that, before Biden's withdrawal, positive sentiment in video posts mentioning Republican candidate is significantly higher than that in video posts mentioning Democratic candidate. However, video posts mentioning Democratic candidate tend to receive more shares and likes compared to video posts mentioning Republican candidate. Following Biden's withdrawal, positive sentiment in video posts mentioning Democratic candidate increases, while negative sentiment decreases. Additionally, the share rate and like rate increase in both video posts mentioning Democratic candidate and video posts mentioning Republican candidate after Biden's withdrawal. Our findings offer valuable insights into the dynamics of sentiments and user engagement rates in online posts mentioning Republican and Democratic candidates during the 2024 \US presidential race.

\section{Ethics Statement}
This research has received ethical approval from the Ethics Review Panel of the University of Luxembourg (ref. ERP 23-053 REMEDIS). All analyses are based on publicly available data. We declare no competing interests.

\begin{acks}
This research is supported by the Luxembourg National Research Fund (FNR) and Belgian National Fund for Scientific Research (FNRS), as part of the project REgulatory Solutions to MitigatE DISinformation (REMEDIS), grant ref. INTER\_FNRS\_21\_16554939\_REMEDIS.
\end{acks}

\end{document}